\documentstyle[floats,epsf,twocolumn,prb,aps]{revtex}

\begin{document}
\draft 

\wideabs{ \title{Electronic Structure of La$_{2-x}$Sr$_x$CuO$_4$ in 
the Vicinity of the Superconductor-Insulator Transition}

\author{A.~Ino$^1$\cite{adr1}, C.~Kim$^2$, M.~Nakamura$^3$, T.~Yoshida$^1$, 
T.~Mizokawa$^1$, Z.-X.~Shen$^2$, A.~Fujimori$^1$,\\ 
T.~Kakeshita$^4$, H.~Eisaki$^4$ and S.~Uchida$^4$}

\address{$^1$ Department of Physics, University of 
Tokyo, Bunkyo-ku, Tokyo 113-0033, Japan} 

\address{$^2$ Department of Applied Physics and Stanford Synchrotron 
Radiation Laboratory, Stanford University,\\ Stanford, CA94305, USA}

\address{$^3$ Department of Physics, Nara University of Education, 
Takabatake-cho, Nara 630-8528, Japan}

\address{$^4$ Department of Superconductivity, University of Tokyo, 
Bunkyo-ku, Tokyo 113-8656, Japan}

\date{\today} 

\maketitle
\begin{abstract}
	We report on the result of angle-resolved photoemission
	(ARPES) study of La$_{2-x}$Sr$_x$CuO$_4$ (LSCO) from an
	optimally doped superconductor ($x=0.15$) to an
	antiferromagnetic insulator ($x=0$).  Near the
	superconductor-insulator transition (SIT) $x\sim 0.05$,
	spectral weight is transferred with hole doping between
	two coexisting components, suggesting a microscopic
	inhomogeneity of the doped-hole distribution.  For the
	underdoped LSCO ($x\leq 0.12$), the dispersive band
	crossing the Fermi level becomes invisible in the
	$(0,0)\!-\!(\pi,\pi)$ direction unlike
	Bi$_2$Sr$_2$CaCu$_2$O$_{8+y}$.  These observations may
	be reconciled with the evolution of holes in the
	insulator into fluctuating stripes in the
	superconductor.
\end{abstract}

}

\narrowtext

The key issue to clarify the nature of high-temperature 
superconductivity in the cuprates is how the electronic structure 
evolves from the antiferromagnetic insulator (AFI) to the 
superconductor (SC) with hole doping.  For the hole-doped CuO$_2$ planes 
in the superconductors, band dispersions and Fermi surfaces have been 
extensively studied by angle-resolved photoemission spectroscopy 
(ARPES) primarily on Bi$_2$Sr$_2$CaCu$_2$O$_{8+y}$ (BSCCO) 
\cite{Shen-rev,Marshall,Ding,FS-LSCO}.  Also for the undoped AFI, band 
dispersions have been observed for Sr$_2$CuO$_2$Cl$_2$ 
\cite{Sr2CuO2Cl2,Kim}.  However, the band structures of the AFI and 
the SC are distinctly different and  ARPES data have been lacking 
around the boundary between the AFI and the SC. In order to reveal the 
missing link, the present ARPES study has been performed on 
La$_{2-x}$Sr$_x$CuO$_4$ (LSCO), which covers continuously from the SC 
to the AFI in a single system.

In addition, the family of LSCO systems show a suppression
of $T_c$ at a hole concentration $\delta\simeq 1/8$, while
the BSCCO system does not.  As the origin of the anomaly at
$\delta\simeq 1/8$, the instability towards the spin-charge
order in a stripe form has been extensively discussed on the
basis of the incommensurate peaks in inelastic neutron
scattering (INS) \cite{Tranquada,Bianconi,Zaanen}. 
Comparing the ARPES spectra of LSCO and BSCCO will help us
to clarify the impact of the stripe fluctuations.

In the present paper, we discuss the novel observation of two spectral 
components coexisting around the SIT ($x \sim 0.05$), the unusual 
disappearance of the Fermi surface near $(\pi/2,\pi/2)$ in the 
underdoped LSCO ($x \le 0.12$) \cite{FS-LSCO}, and their relevance to 
the stripe fluctuations.

\begin{figure*}
	\centerline{\epsfxsize=133mm \epsfbox{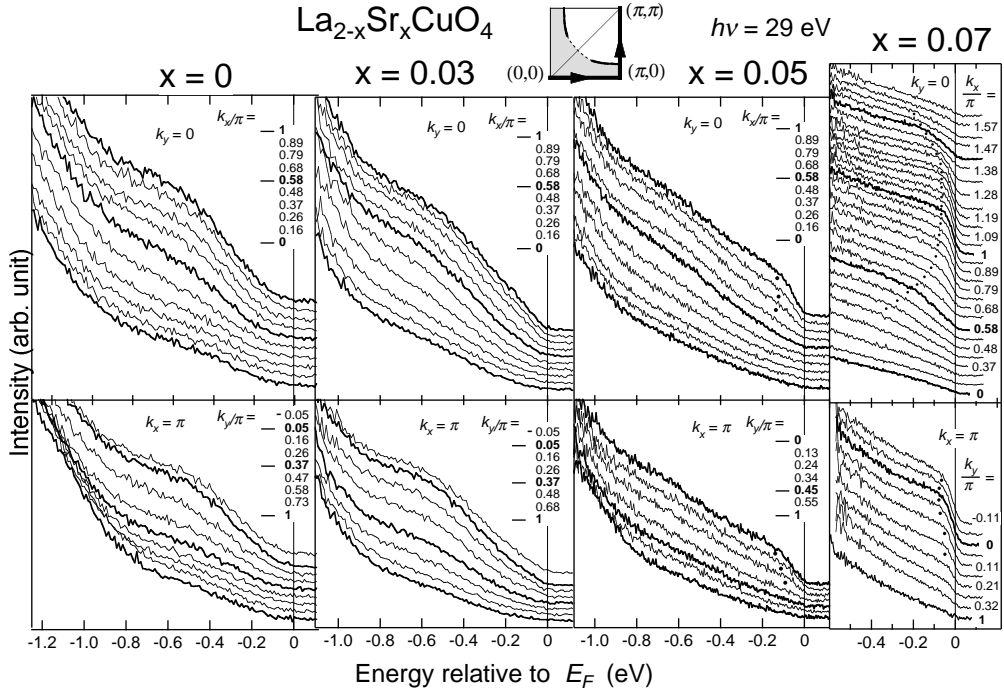}} \vspace{0.5pc}
	\caption{ARPES spectra of La$_{2-x}$Sr$_x$CuO$_4$ (LSCO)
	for $x=0$, 0.03, 0.05 and 0.07, taken along
	$(0,0)\rightarrow(\pi,0)$ (upper panels) and
	$(\pi,0)\rightarrow(\pi,\pi)$ (lower panels).  The
	dispersive component emerging around $\vec{k} = (\pi,0)$
	is sufficiently strong compared to the angle-independent
	background.  Inset shows the Brillouin zone and the
	Fermi surface of underdoped LSCO\protect\cite{FS-LSCO}.}
	\label{00pi0pipi}
\end{figure*}

\begin{figure}[!t]
	\epsfxsize=90mm \centerline{\epsfbox{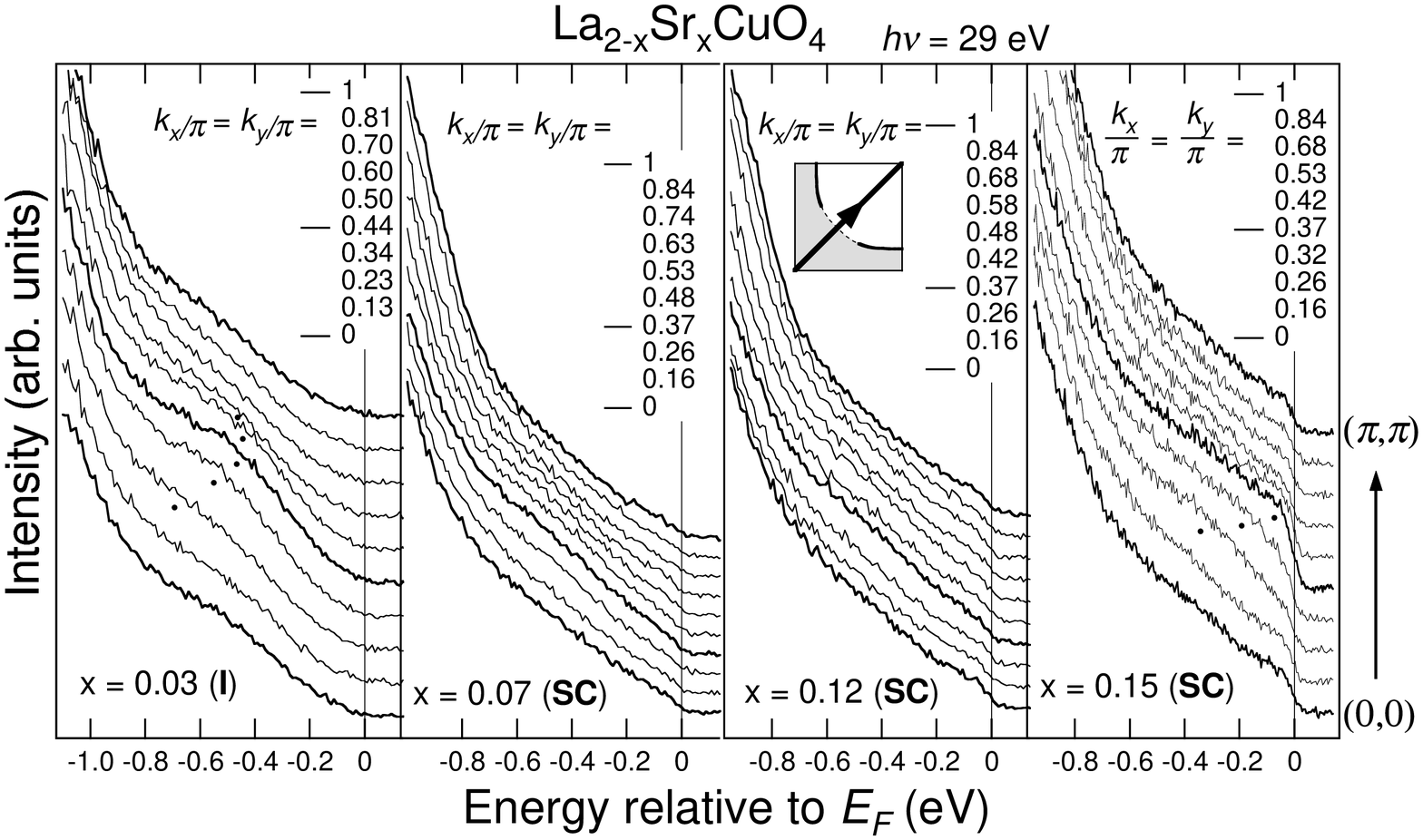}} 
	\vspace{0.5pc} \caption{Doping dependence of ARPES spectra taken 
	along (0,0)$\rightarrow$($\pi$,$\pi$).  While a broad feature 
	dispersing across $E_F$ is identified for $x=0.15$, the dispersive 
	band is absent near $E_F$ for $x=0.12$ and 0.07, even though the 
	system is superconducting.  For $x=0.03$, a band appears at 
	$\sim-0.45$ eV around ($\pi/2$,$\pi/2$).}
	\label{00pipi}
\end{figure}

The ARPES measurements were carried out at beamline 5-3 of
Stanford Synchrotron Radiation Laboratory (SSRL).  Incident
photons had an energy of 29 eV and were linearly polarized. 
The total energy resolution was approximately 45 meV and the
angular resolution was $\pm 1$ degree.  Single crystals of
LSCO were grown by the traveling-solvent floating-zone
method and then annealed so that the oxygen content became
stoichiometric.  The accuracy of the hole concentration
$\delta$ was $\pm0.01$.  The $x=0$ samples were slightly
hole doped by excess oxygen so that $\delta\simeq0.005$ was
deduced from its N\'{e}el temperature $T_N = 220$ K
\cite{CYChen}.  The spectrometer was kept in an ultra high
vacuum better than $5\times 10^{-11}$ Torr during the
measurements.  The samples were cleaved {\it in situ} by
hitting a post glued on the top of the samples.  The
measurements were done only at low temperatures ($T \sim 11$
K), because the surfaces degraded rapidly at higher
temperatures.  Throughout the paper, the spectral intensity
for different angles are normalized to the intensity of the
incident light.  In the analysis, the spectrum at (0,0) is
assumed to represent the angle-independent background,
because emission from the states of $d_{x^2-y^2}$ symmetry
is not allowed for the direction normal to the CuO$_2$ plane
due to a selection rule.  Indeed, Figs.~\ref{00pi0pipi} and
\ref{00pipi} show that spectra in the vicinity of the Fermi
level ($E_F$) are angle-independent, when there are no
dispersive features.


\begin{figure}[!t]
	\centerline{\epsfxsize=68mm \epsfbox{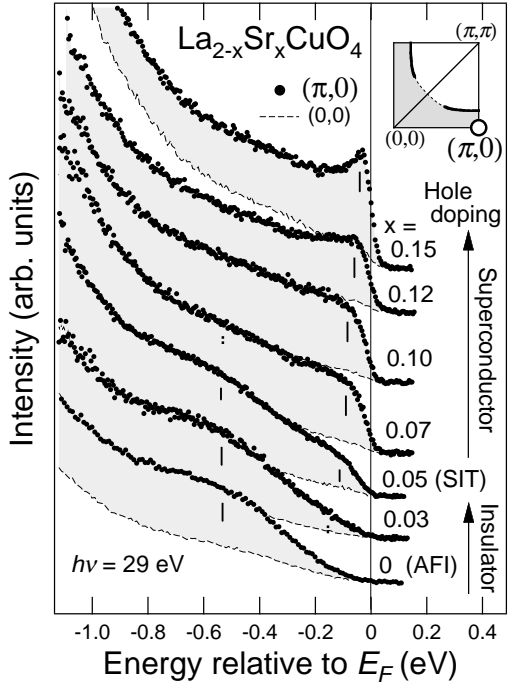}} \vspace{0.5pc}
	\caption{Doping dependence of ARPES spectra at $\vec{k}
	= (\pi,0)$.  The spectra have been normalized to the
	intensity of the valence-band peak at $\sim -3$ eV. Each
	thin dashed line denotes the spectrum at (0,0)
	representing the angle-independent background.  The
	spectrum of $x=0.05$ at the superconductor-insulator
	transition (SIT) point shows two spectral features
	associated with the antiferromagnetic insulator (AFI)
	and the superconductor (SC) at $\sim -0.5$ eV and $\sim
	-0.1$ eV, respectively.}
	\label{pi0}
\end{figure}

ARPES spectra along
$(0,0)\rightarrow(\pi,0)\rightarrow(\pi,\pi)$ clearly shows
angle dependence as shown in Fig.~\ref{00pi0pipi}.  Even
though spectral features are broad for underdoped and
heavily undoped
cuprates\cite{Shen-rev,Marshall,Sr2CuO2Cl2,Kim}, the
dispersive component emerging around $\vec{k} = (\pi,0)$ is
sufficiently strong compared to the angle-independent
background.  Figure~\ref{pi0} shows the doping dependence of
the ARPES spectrum at $(\pi,0)$.  As reported previously
\cite{FS-LSCO}, a relatively sharp peak is present just
below $E_F$ for the optimally doped sample ($x=0.15$).  For
the underdoped samples ($x=0.12, 0.10$ and 0.07), the peak
is broadened and shifted downwards.  When the hole
concentration is further decreased to the vicinity of the
SIT ($x=0.05$), the peak near $E_F$ rapidly loses its
intensity and concomitantly another broad feature appears
around $-0.55$ eV. In the AFI phase ($x=0$), the peak near
$E_F$ disappears entirely while the structure around
$\sim-0.55$ eV becomes predominant.  As for LSCO,
Fig.~\ref{pi0} is different from the scenario that a single
peak is shifted downwards and continuously evolves from the
SC into the AFI as seen in BSCCO and Ca$_2$CuO$_2$Cl$_2$
\cite{Filip}, and rather indicates that the spectral weight
is transferred between the two features originated from the
SC and the AFI, and

On the other hand, ARPES spectra in the (0,0)-($\pi$,$\pi$)
direction show a different doping dependence.  The spectra
for representative doping levels are shown in
Fig.~\ref{00pipi}.  For the optimally doped sample
($x=0.15$), one can identify a band crossing $E_F$ around
(0.4$\pi$,0.4$\pi$), although the dispersive feature is
considerably weak.  For the underdoped samples ($x=0.12,
0.10$ and 0.07), the band crossing $E_F$ disappears, even
though the system is still superconducting.  The invisible
band crossing along (0,0)-($\pi$,$\pi$) is reproduced for
several samples for $0.07 \le x \le 0.12$, excluding
accidentally inferior surfaces as its origin.  For the
insulating sample ($x=0.03$ and 0), a broad feature appears
around $\sim -0.45$ eV near ($\pi/2$,$\pi/2$), correlated
with the growth of the broad structure around $\sim-0.55$ eV
at ($\pi$,0).

\begin{figure}[!t]
	\epsfxsize=87mm \centerline{\epsfbox{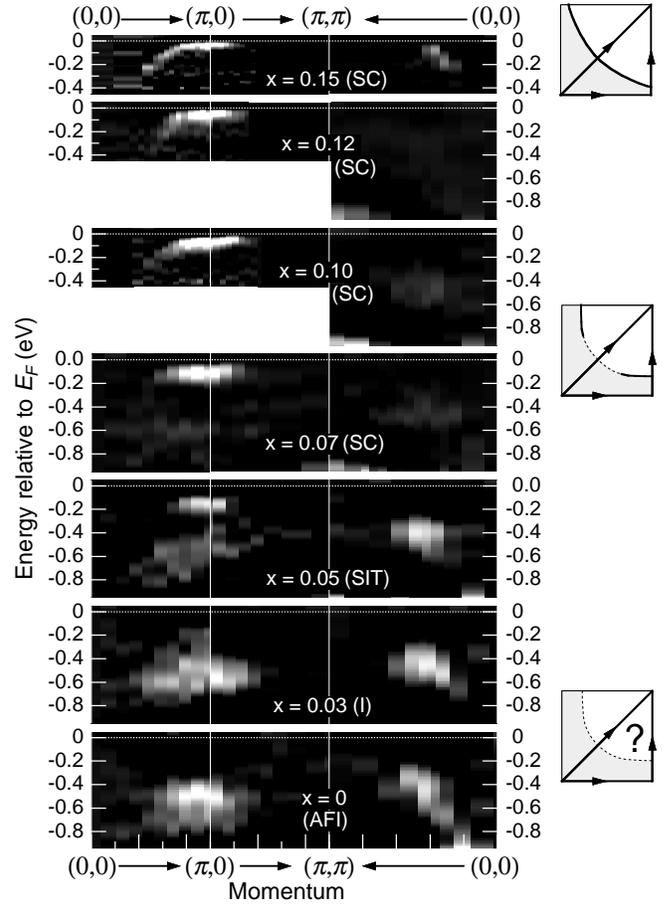}} \vspace{0.5pc} 
	\caption{Band structure deduced from the ARPES spectra by taking 
	the second derivatives after subtracting the angle-independent 
	background.  For $x=0.05$, while the band around $-0.1$ eV 
	originates from the band in the underdoped superconductor 
	($x=0.12$, 0.10 and 0.07), the band around $-0.5$ eV appears to 
	originate from the lower Hubbard band in the antiferromagnetic 
	insulator ($x=0$).}
	\label{map}
\end{figure}


Overall dispersions of the spectral features have been
derived from the ARPES spectra by taking second derivatives
as shown in Fig.~\ref{map}.  The band near $E_F$ for
$x=0.05$, 0.07, 0.10 and 0.12 has a dispersion similar to
that for $x=0.15$ around $(\pi,0)$: when one goes as
$(0,0)\!\to\!(\pi,0)\!\to\!(\pi,\pi)$, the band approaches
$E_F$ until $\sim$(0.8$\pi$,0), stays there until ($\pi$,0),
then further approaches $E_F$ and goes above $E_F$ through
the superconducting gap around $\sim$$(\pi,\pi/4)$.  The
band seen near $E_F$ should be responsible for the
superconductivity.  On the other hand, the dispersions of
the broad feature seen around $-0.5$ eV are almost the same
among $x=0$, 0.03 and 0.05 and similar to the band
dispersion of the undoped CuO$_2$ plane in
Sr$_2$CuO$_2$Cl$_2$ \cite{Sr2CuO2Cl2,Kim} and
PrBa$_2$Cu$_3$O$_7$ \cite{PBCO}.  Along the
$(0,0)$$\to$$(\pi,\pi)$ cut, the broad peak moves upwards,
reaches a band maximum ($\sim-0.45$ eV) around
$(\pi/2,\pi/2)$ and then disappears.  The broad peak emerges
in going from $(0,0)$ to $(\pi,0)$ and then disappears
between $(0,0)$ and $(\pi,\pi)$.  Therefore, the band around
$-0.5$ eV originates from the lower Hubbard band (LHB) of
the AFI.

\begin{figure*}[!t]
	\epsfxsize=175mm\centerline{\epsfbox{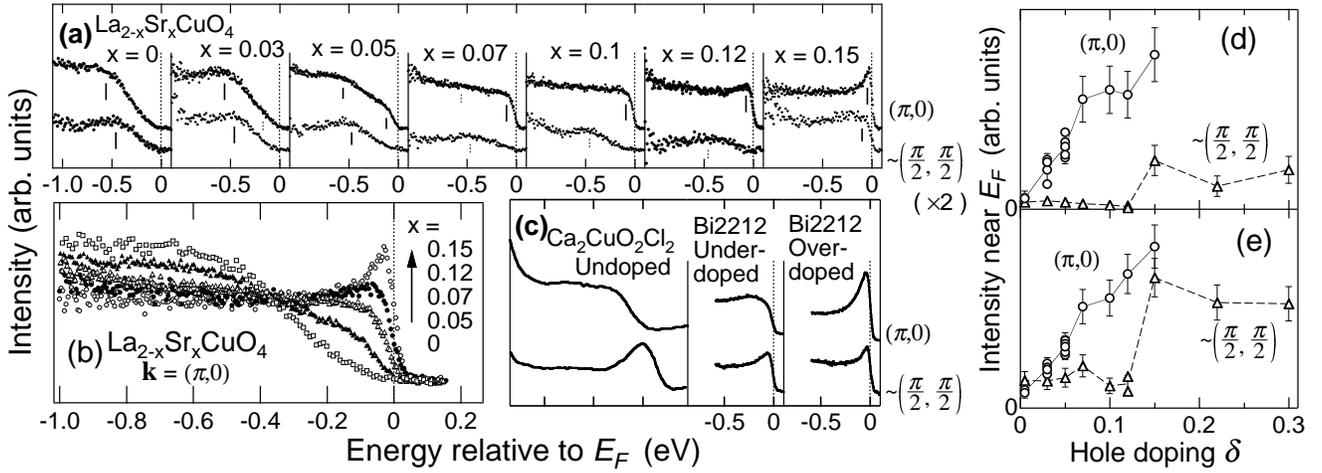}}\vspace{0.5pc}
	\caption{(a) Comparison between the spectra at ($\pi$,0)
	and $\sim$($\pi/2$,$\pi/2$) (multiplied by 2) for
	La$_{2-x}$Sr$_x$CuO$_4$.  The spectrum at (0,0) has been
	subtracted as the angle-independent background.  (b)
	Spectra at ($\pi$,0), normalized to the integrated
	intensity in $E > -0.9$ eV, clearly demonstrating the
	systematic evolution with hole doping.  (c) ARPES
	spectra for other cuprates: Ca$_2$CuOCl$_2$ (undoped)
	and Bi$_2$Sr$_2$CaCu$_2$O$_{8+y}$ (underdoped and
	overdoped), shown in a similar way to (a), taken from
	Ref.~\protect\onlinecite{Kim} and
	\protect\onlinecite{Filip}, respectively.  While the
	lineshapes at ($\pi$,0) are similar between
	La$_{2-x}$Sr$_x$CuO$_4$ and
	Bi$_2$Sr$_2$CaCu$_2$O$_{8+y}$, the lineshapes at
	$\sim$($\pi/2$,$\pi/2$) are quite different.  (d)(e)
	Doping dependence of the spectral intensity near $E_F$
	($E>-0.2$ eV) at ($\pi$,0) ($\bigcirc$) and at
	$\sim$($\pi/2$,$\pi/2$) ($\bigtriangleup$), normalized
	to the intensity of the valence-band peak of the
	($\pi$,0) spectra at $\sim-3$ eV and to the integrated
	intensity in $E > -0.9$ eV for (d) and (e),
	respectively.}
	\label{int}
\end{figure*}
\begin{figure}[!t]
	\epsfxsize=85mm \centerline{\epsfbox{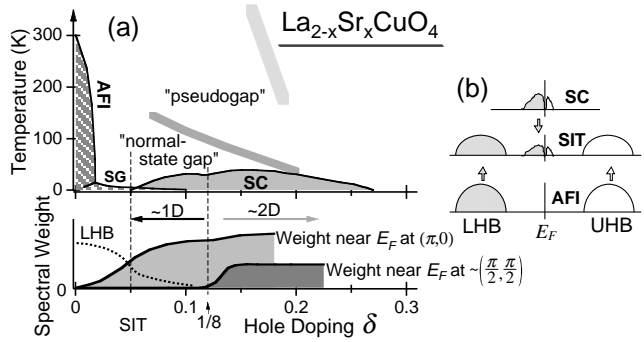}} 
	\vspace{0.5pc}%
	\caption{(a) Schematic picture of the evolution of spectral weight 
	with hole doping for the lower Hubbard band (LHB) and the band 
	near $E_F$ at $(\pi,0)$ and at $\sim(\pi/2,\pi/2)$.  The phase 
	diagram was drawn after Refs.~\protect\onlinecite{muSR} and 
	\protect\onlinecite{Momono} (SG: so-called spin-glass phase).  (b) 
	Schematic drawing indicating that the spectra around the SIT 
	($x\sim0.05$) consist of two kinds of electronic states derived 
	from the SC and the AFI.}
	\label{schematic}
\end{figure}

In Fig.~\ref{int}(a), the dispersive components of the ARPES
spectra are compared between ($\pi$,0) and
$\sim$($\pi/2$,$\pi/2$).  Around the SIT ($x\sim0.05$), the
two spectral features coexist at ($\pi$,0), while only one
broad peak is observed at $\sim$($\pi/2$,$\pi/2$).  This
excludes extrinsic origins for the two structures such as a
partial charge-up of the sample.  Figure~\ref{int}(b)
demonstrates that the spectral lineshape at ($\pi$,0) and
the relative intensity of two structures are quite
systematically evolves with hole doping, indicating that
surfaces of good quality were consistently obtained for all
the doping levels.  The spectral weight transfer is
reminiscent of earlier discussion based on angle-integrated
data of LSCO and Nd$_{2-x}$Ce$_x$CuO$_4$\cite{Allen}.  A
possible origin for the coexistence of the two spectral
features is a phase separation into hole-poor
antiferromagnetic (AF) domains and hole-rich superconducting
domains.  The spectra around SIT may be regarded as a
superposition of the spectra of the SC and the AFI, as
illustrated in Fig.~\ref{schematic} (b).  Indeed, when holes
are doped into La$_2$CuO$_{4+y}$ with excess oxygens, such a
phase separation occurs macroscopically as revealed by,
e.g., neutron diffraction \cite{Radaelli}, but corresponding
observation has never been reported for the Sr-doped LSCO
system.  A more likely interpretation is a microscopic
inhomogeneity of the hole density in the sense that the
doped holes are segregated in boundaries of AF domains on
the scale of several atomic distances.  Indeed, $\mu^{+}$SR
\cite{muSR} and $^{139}$La NQR \cite{NQR} experiments have
shown the presence of a local magnetic field in the
so-called spin-glass phase [Fig.~\ref{schematic} (a)]. 
Then, since the present spectra were taken above the
spin-glass transition temperature, splitting into the two
structures would be due to dynamical fluctuations of such a
microscopic phase separation.  Furthermore, the microscopic
phase separation may explain why the chemical potential is
pinned against hole doping for $x\lesssim0.12$ \cite{chem}. 
However, as for the underdoped BSCCO, the splitting of the
two components by $\sim0.5$ eV has not been reported so far
and the peak at ($\pi$,0) seems to be shifted smoothly in
going from the SC to the AFI \cite{Filip}.  The difference
might imply that the tendency toward the hole segregation is
stronger in LSCO than in BSCCO.

The ARPES spectra of LSCO are compared with those of
BSCCO\cite{Kim,Shen&Schrieffer} in Figs.~\ref{int} (a) and
(c).  Whereas the lineshapes at ($\pi$,0) are similar
between LSCO and BSCCO irrespective of doping levels, the
spectra near ($\pi/2$,$\pi/2$) are quite different: while
the peak near $E_F$ is sharp for both the overdoped and
underdoped BSCCO, one finds no well-defined feature around
$E_F$ for underdoped LSCO. This difference is likely to be
related with the stripe fluctuations, which have more static
tendency in LSCO than in BSCCO, judging from the suppression
of $T_{c}$ at $\delta\simeq1/8$.  Also for the BSCCO system,
it has been reported that the sharp peak near $E_F$ is
suppressed near $(\pi/2,\pi/2)$ upon Zn-doping, which is
considered to pin the dynamical stripe correlations
\cite{Zn-BSCCO,Akoshima}.  The absence of the band crossing
$E_F$ near $(\pi/2,\pi/2)$ may be reconciled with the
vertically and horizontally oriented stripes in LSCO
\cite{Tranquada}.  Intuitively, while the system may be
metallic along the half-filled stripes, namely, in the
$(0,0)$-$(\pi,0)$ or $(0,0)$-$(0,\pi)$ direction, the
low-energy excitations should be strongly suppressed in the
directions crossing all the stripes such as the
$(0,0)$-$(\pi,\pi)$ direction.  This conjecture was
supported by a recent numerical study of small clusters with
charge stripes \cite{Tohyama}, and is consistent with the
suppression of the Hall coefficient in the stripe-ordered
state of La$_{1.6-x}$Nd$_{0.4}$Sr$_x$CuO$_4$ for $x<1/8$
\cite{Noda}.

The doping dependence of the spectral intensity near $E_{F}$
($E>-0.2$ eV) is shown in Figs.~\ref{int} (d) and (e) for
two normalization methods.  Note that the essential features
are independent of normalization.  Our picture of the
evolution of the spectral weight is schematically drawn in
Fig.~\ref{schematic}.  In the ARPES spectra, the intensity
near $E_F$ appears at ($\pi$,0) with hole doping for
$x\gtrsim0.05$, where the incommensurability of the spin
fluctuations also arises according to the INS study
\cite{Yamada}.  On the other hand, the intensity near
($\pi/2$,$\pi/2$) remains suppressed with hole doping for
the entire underdoped region ($0.05 \lesssim x \lesssim
0.12$).  Hence, one may think that the segregated holes for
$x\sim0.05$ already start to be arranged vertically and
horizontally.  Therefore we propose that the hole-rich
boundaries of the AF domains around the SIT continuously
evolves into the stripe correlations in the underdoped SC.
In going from $x=0.12$ to 0.15, the Fermi-surface crossing
appears in the (0,0)-($\pi$,$\pi$) direction, probably
corresponding to the phenomenon that the incommensurability
in INS saturates for $x\gtrsim0.15$ \cite{Yamada}.  This may
be understood that the doped holes in excess of $x=1/8$
overflow the saturated stripes and that the two-dimensional
electronic structure recovers.

In conclusion, we have shown that the SC and AFI characters 
coexist in the ARPES spectra in the vicinity of the SIT for 
LSCO. The band crossing $E_F$ disappears near 
($\pi/2$,$\pi/2$) for the underdoped LSCO, associated with 
the formation of the dynamical stripes.  The present 
observations provide a new perspective of how the holes 
doped in the AFI evolve into the fluctuating stripes in the 
underdoped SC. Mechanism in which the SC to AFI transition 
occurs is a subject of strong theoretical interest 
\cite{SO5,Assaad&Imada} and should be addressed by further 
studies.

We would like to thank T.~Tohyama and 
S.~Maekawa for enlightening discussions.  This work was supported by 
the New Energy and Industrial Technology Development Organization 
(NEDO), Special Coordination Fund for Promoting Science and Technology 
from the Science and Technology Agency of Japan, the U.~S.~DOE, Office 
of Basic Energy Science and Division of Material Science.  Stanford 
Synchrotron Radiation Laboratory is operated by the U.~S.~DOE, Office 
of Basic Energy Sciences, Division of Chemical Sciences.

\end{document}